\documentclass{ws-procs961x669}            
\usepackage{graphicx}	

\begin{document}
\title{Hints of the $H_0-r_d$ tension in uncorrelated Baryon Acoustic Oscillations dataset}

\author{Denitsa Staicova}

\address{Institute for Nuclear Research and Nuclear Energy, \\
Bulgarian Academy of Sciences, Sofia, Bulgaria\\
E-mail: dstaicova@inrne.bas.bg}

\begin{abstract}
Baryon Acoustic Oscillations (BAO) datasets use very precise measurements of the spatial distribution of large-scale structures as a distance ladder to help constrain cosmological parameters. In a recent article \cite{Benisty:2020otr}, we combined 17 uncorrelated BAO measurements in the effective redshift range $0.106 \le z \le 2.36$ with the Cosmic Chronometers data, the Pantheon Type Ia supernova and the Hubble Diagram of Gamma Ray Bursts and Quasars to obtain that the $\Lambda$CDM model fit infers for the Hubble constant: $69.85 \pm 1.27km/sec/Mpc$ and for the sound horizon distance: $146.1 \pm 2.15Mpc$. Beyond the $\Lambda$CDM model we test $\Omega_k$CDM and wCDM and we get $\Omega_k = -0.076 \pm 0.012$,  $w = -0.989 \pm 0.049$  accordingly. In this proceeding we present elaborate on our findings and we compare them to other recent results in the literature.
\end{abstract}


\bodymatter

\section{Introduction}\label{aba:sec1}
The $\sim 4\sigma$ tension between the Hubble parameter measured by late universe observations by the SH0ES collaboration  \citep{Riess:2019cxk} and the one measured from the cosmic microwave background (CMB) by the Planck Collaboration (\cite{Aghanim:2018eyx}) is one of the major stumbling block in front of modern cosmology and the theories aiming to explain the evolution of the Universe (\cite{DiValentino:2020vhf,DiValentino:2020zio,DiValentino:2020vvd,Efstathiou:2020wxn,Borhanian:2020vyr,Hryczuk:2020jhi,Klypin:2020tud,Ivanov:2020mfr,Chudaykin:2020acu,Lyu:2020lps,Alestas:2020mvb,Motloch:2019gux,Frusciante:2019puu,Staicova:2020wph, Staicova:2019ksr, Benisty:2021sul, AresteSalo:2021lmp,Benisty:2020kdt,Bahamonde:2021gfp}). The default $\Lambda$CDM model which uses a combination between cold dark matter and dark energy components has been shown to fit remarkably well current astronomical observations yet it fails to explain not only the beginning of the universe (the inflationary epoch) but also the Hubble tension and the related $\sigma_8$ tension.

In a recent article \cite{Benisty:2020otr} we selected 17 uncorrelated BAO points from the largest collection of BAO data points (333 points). We then combined them with the Cosmic Chronometers data, the Pantheon Type Ia supernova, and the Hubble Diagram of Gamma-Ray Bursts and Quasars. From this combination of datasets, referred in the article sometimes as the  {\em full} dataset, we found: the Hubble constant yields $ 69.85 \pm 1.27 km/sec/Mpc$, the sound horizon distance is $ 146.1 \pm 2.15 Mpc$ and the matter energy density -- $\Omega_m=0.271 \pm 0.016$. If one uses the so called {\em Riess prior} (denoted here as R19) to constrain $H_0$ by the model-independent local universe measurement \cite{Riess:2019cxk}, one gets: $H_0=71.40 \pm0.89$, $r_d=143.5 \pm2.0$ and $\Omega_m=0.267 \pm0.017$. Beyond the $\Lambda$CDM model we test $\Omega_K$CDM and wCDM. The spatial curvature is $\Omega_k = -0.076 \pm 0.012$ and the dark energy equation of states is $w = -0.989 \pm 0.049$. In this proceeding we discuss how our results are situated regarding other published results by emphasizing on the need to consider the $H_0$-tension in the context of the $H_0-r_d$ plane or even of the $H_0-r_d-\Omega_m$ plane.

\section{Overview of the used datasets}

The Baryon acoustic oscillations (BAO) are fluctuations in the photon-baryonic plasma that froze at recombination at the so called drag sound horizon. Because the sound horizon can be calculated rather simply from basic assumptions of the pre-recombination plasma, they provide a standard ruler which can be seen in the clustering of large scale structures. This provides an independent way to probe cosmological parameters, complimentary to this of the Supernova and the CMB surveys, see (\cite{Handley:2019tkm,DiValentino:2020hov,Luo:2020ufj}). The BAO peak can be measured from objects with different nature and using different methods. For example: the BOSS experiment measures the clustering of different galaxies: emission-line galaxies (ELGs),  luminous red galaxies (LRGs), and quasars, and also from the correlation function of the Lyman-alpha (Ly$\alpha$) absorption lines in the spectra of distant quasars etc. The peak can be seen on different redshifts, providing us with a standard ruler evolving with the Universe since the recombination epoch (\cite{Cuceu:2019for,Wu:2020nxz}).

The final dataset we use is a set of uncorrelated data points from different \textbf{ BAO} measurements: the Sloan Digital Sky Survey (SDSS), the WiggleZ Dark Energy Survey, Dark Energy Camera Legacy Survey (DECaLS), the Dark Energy Survey (DES), the 6dF Galaxy Survey (6dFGS)
(\cite{Beutler:2011hx, Ross:2014qpa,Percival:2009xn,Tojeiro:2014eea,Blake:2012pj,Seo:2012xy,Anderson:2012sa,Sridhar:2020czy,Bautista:2017wwp,Abbott:2017wcz,Hou:2020rse,Ata:2017dya,Busca:2012bu,Agathe:2019vsu, Carter:2018vce, Kazin:2014qga}). To this dataset, we add \textbf{cosmic chronometers (CCs)} (30 uncorrelated CC measurements of $H(z)$ \cite{Moresco:2012by,Moresco:2012jh,Moresco:2015cya,Moresco:2016mzx}), and \textbf{standard candles (SCs)} (the Pantheon Type Ia supernova dataset \cite{Perlmutter:1998np,Riess:1998cb, Scolnic:2017caz, Anagnostopoulos:2020ctz}), and  quasars \cite{Roberts:2017nkm} and gamma-ray bursts \cite{Demianski:2016zxi} (186 points).

\section{Theoretical background}
We use the following theoretical setup. If one assumes a Friedmann-{Lema\^itre}-Robertson-Walker metric with the scale parameter $a = 1/(1+z)$, where $z$ is the redshift, one gets for the Friedmann equation for the $\Lambda$CDM model:
\begin{equation}
    E(z)^2 = \Omega_{r} (1+z)^4 + \Omega_{m} (1+z)^3 + \Omega_{k} (1+z)^2 + \Omega_{\Lambda},
    \label{eq:hzlcdm}
\end{equation}
where $\Omega_{r}$, $\Omega_{m}$, $\Omega_{\Lambda}$, and $\Omega_{k}$ are respectively the fractional densities of radiation, matter, dark energy, and the spatial curvature at redshift $z=0$. Here $E(z)=H(z)/H_0$, and $H(z) = \dot{a}/a$ is the Hubble parameter at $z$, while $H_0$ is the Hubble parameter today. The radiation density  can be computed as $\Omega_r = 1 - \Omega_m - \Omega_\Lambda - \Omega_{k}$. For wCDM the Friedmann equation is generalized to $\Omega_{\Lambda}\to \Omega_{DE}^{0}  (1+z)^{-3(1+w)}$, while $\Omega_k=0$ represents a flat universe.

Since in cosmology one deals with the measurements of angles and redshifts, it is needed to connect the different cosmological distances with the observational quantities. The comoving angular diameter distance: (\cite{Hogg:2020ktc,Martinelli:2020hud})
\begin{equation}
D_M=\frac{c}{H_0} S_k\left(\int_0^z\frac{dz'}{E(z')}\right),
\end{equation}
where one accounts for non-zero spatial curvature with:
 {\begin{equation}
S_k(x) =
\begin{cases}
\frac{1}{\sqrt{\Omega_k}}\sinh\left(\sqrt{\Omega_k}x\right) \quad &\text{if}\quad \Omega_k>0
\\
x \quad  &\text{if} \quad \Omega_k=0
\\
\frac{1}{\sqrt{-\Omega_k}}\sin\left(\sqrt{-\Omega_k} x\right)\quad &\text{if} \quad \Omega_k<0
\end{cases}.
\end{equation}}

The other distances we use are the Hubble distance  $D_H(z)= c/H(z)$, the angular diameter distance $D_A=D_M/(1+z)$ and the volume averaged distance:
\begin{equation}
    D_V(z) \equiv [ z D_H(z) D_M^2(z) ]^{1/3}.
\end{equation}

As said before the BAO use as a standard ruler depends only on the sound horizon $r_d$ at the drag epoch  ({$z_d \approx 1060$}) when photons and baryons decouple:
\begin{equation}
r_d = \int_{z_d}^{\infty} \frac{c_s(z)}{H(z)} dz
,\end{equation}
where $c_s \approx c \left(3 + 9\rho_b /(4\rho_\gamma) \right)^{-0.5}$ is the speed of sound in the baryon-photon fluid with the baryon $\rho_b(z)$ and the photon $\rho_\gamma(z)$ densities, respectively (\cite{Aubourg:2014yra}). One needs to acknowledge that since the actual measured quantities are the projections  $\Delta z= r_d H/c$ and $\Delta \theta=r_d /(1+z) D_A(z)$, where $\Delta z$ and $\Delta \theta$ are the redshift and the angular separation, from BAO one can get information only about the quantity $r_d \times H$. Thus in order to decouple these quantities, one needs some kind of independent measurement of $H_0$ or an assumption for $r_d$. Here we take $r_d$ as an independent parameter but use the additional datasets (CC+SC) to decouple the two variables. In a recent paper \cite{Benisty:2021gde}, we instead use the combination $r_d \times H_0$ as a parameter to break the degeneracy.

\section{Numerical methods}
To deal with the possible correlations in the BAO dataset, we perform a covariance analysis based on the one proposed in Ref. \cite{Kazantzidis:2018rnb}. We transform the  standard covariance matrix for uncorrelated points $C_{ii}$ into $C_{ij}$ as follows:
\begin{equation}
C_{ii} = \sigma_i^2 \quad \longrightarrow \quad C_{ij} = \sigma_i^2+0.5 \sigma_i \sigma_j
\end{equation}
by adding randomly certain number of nondiagonal elements  while keeping it symmetric. Here $\sigma_i \sigma_j$ are the published $1\sigma$ errors of the data points $i,j$. With this approach we show that the effect of  up to $25\%$ random correlations with this magnitude results in less than $10\%$ deviation in the final results, thus it is minimal and the points can be considered uncorrelated.

We use a nested sampler as  implemented within the open-source package $Polychord$ (\cite{Handley:2015fda}) with the $GetDist$ package (\cite{Lewis:2019xzd}) to present the results. The priors we use to obtain the results and the averaged mean values across the 3 models ($\Lambda$CDM, $w$CDM, $\Omega_k$CDM)  can be found in Table \ref{tbl1}.

\section{Comparison with other results}

The complete numerical results for the three models and the three datasets (BAO+R19, BAO+SC+CC, BAO+SC+CC+R19) can be found in Table 3 in \cite{Benisty:2020otr}. In this article we will focus on comparing our results with other works, by using only the {\em full} datasets with and without the R19 prior, referred to as BAO+SC+CC+R19 and BAO+SC+CC.

\begin{figure}[!h]%
\centering
  \includegraphics[ width=0.495\textwidth]{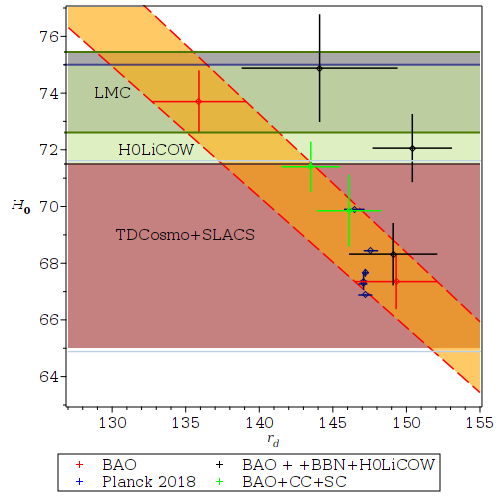}
  \includegraphics[ width=0.495\textwidth]{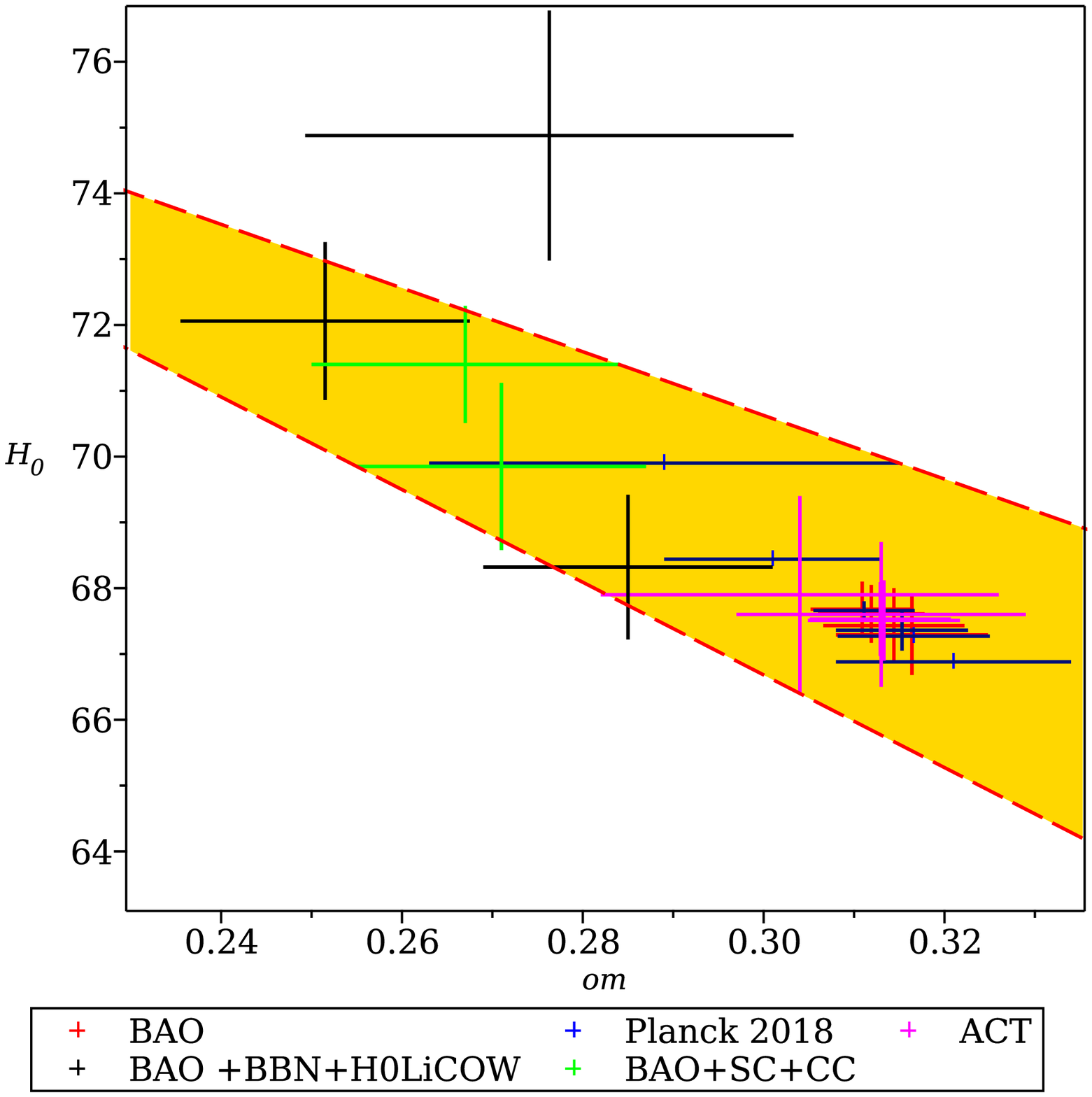}
  \caption{a) $H_0-r_d$ plane comparison of different results. b) $H_0-\Omega_m$ plane comparision. Our points are with green, for legend see the text }%
  \label{fig1}
\end{figure}

Specifically, on Fig. \ref{fig1}, we use some known measurements in the $H_0-r_d$ and $H_0-\Omega_m$ plane. It has been discussed in a number of papers, more notably \cite{Knox_2020},  that the $H_0$-tension is actually extended to a tension with respect to the  main parameters $H_0$, $r_d$ and $\Omega_m$, since increasing $\Omega_m$ to reduce $r_d$ leads to the opposite to the desired effect on $H_0$. To demonstrate this, we have plotted the following points: TDCosmo IV \cite{Birrer:2020tax}, 
H0LiCOW XIII \cite{Wong:2019kwg}, 
LMC Cepheids \cite{Riess:2019cxk}, 
Planck 2018 \cite{Aghanim:2018eyx}, 
eBOSS SDSS-IV \cite{eBOSS:2020yzd}, 
BAO +BBN+H0LiCOW and BAO +BBN+CC \cite{Nunes:2020uex}, all listed for convenience in Table \ref{tbl2}. Note, some references do not measure $r_d$, others omit mentioning $\Omega_m$.

On Fig. 1 a) we consider only $H_0-r_d$ and we see that our results (in green) fit nicely between the results of the pure BAO points by eBOSS (in red) and the Planck 2018 points (in blue).

On this plot, the points we obtain with the R19 prior (top green point) are very close to the local universe measurements (LMC), while the points we obtain with a uniform prior (the lower green point) are close to the early universe (i.e. Planck 2018 with bue). The numbers we obtain are surprisingly close to a measurement of the Tip of the Red Giants Branch $H_0=69.8 \pm 1.9 (km/s)/Mpc$ \cite{Freedman:2020dne} - an independent measurement based on the distances to nearby galaxies (in this case based on TRGB stars in the Large Magellanic Cloud, grounded on detached eclipsing binaries).

As for $r_d$, we see that our points without the R19 prior are again close to the Planck measurement (Planck 2018 $r_d = 147.09 \pm 0.26 Mpc$, ours: $r_d =146.1\pm 2.2 Mpc$). The final measurements from the completed SDSS lineage of experiments in large-scale structure provide $r_d = 149.3 \pm 2.8 Mpc$ (\cite{Alam:2020sor}). Using BAO, SNea, the ages of early-type galaxies, and local determinations of the Hubble constant,   Ref. \cite{Verde:2016ccp} reports $r_d = 143.9\pm3.1 $ Mpc.  Thus, one can see clearly the tension between different results from the early and late universe confirming the ``tensions in the $r_{d}\,-\,H_0$ plane.''  \cite{Knox_2020}. The model with the R19 prior gives $r_d = 143.5 \pm 2.0 Mpc$. Importantly,  the choice of a prior for $r_d$ has a critical effect on $H_0$ decreasing or increasing the inferred value to a large extend independently of the prior on $H_0$.

Another part of the $H_0-r_d$ puzzle is the effect of the matter energy density $\Omega_m$. A comparison of different results for $\Omega_m$ can be seen on Fig. 1 b), where we have plotted the $H_0-\Omega_m$ plane. Note that the eBOSS points used here are different from the ones used in Table 2, as we use different set of published points (Table 4 in Ref. \cite{eBOSS:2020yzd}). Also, we add the points by ACT which can be found in Table 4 in Ref. \cite{ACT:2020gnv}. Our points are once again in green and as we can see they fit very close to Plancks's results in blue which come with very small error on $H_0$ but rather large on $\Omega_m$ ($\Omega_m^{Planck}=.289,0\pm 0.03$). In our case, the matter density is lower than the expected. One can also note that removing the top right point  (belonging to BAO +BBN+H0LiCOW, see Table 2) will make the points along with the error bars to lie on approximately one plane, similar to the one in the $H_0-r_d$ plane and thus hinting at possible degeneracy (in orange and yellow respectively). This has been commented already in \cite{Knox_2020} and is related to the fact we do not know how exactly the phase space of all parameters looks like. The tension for $H_0-r_d-\Omega_m$ appears in different measurements (Planck, BAO, ACT etc), some from the early Universe, others from the late one and the independent local measurements. This once again poses the question whether we need to consider the full parameter space when discussing the tension.

\begin{figure}[!ht]%
\centering
  \includegraphics[width=0.5\textwidth]{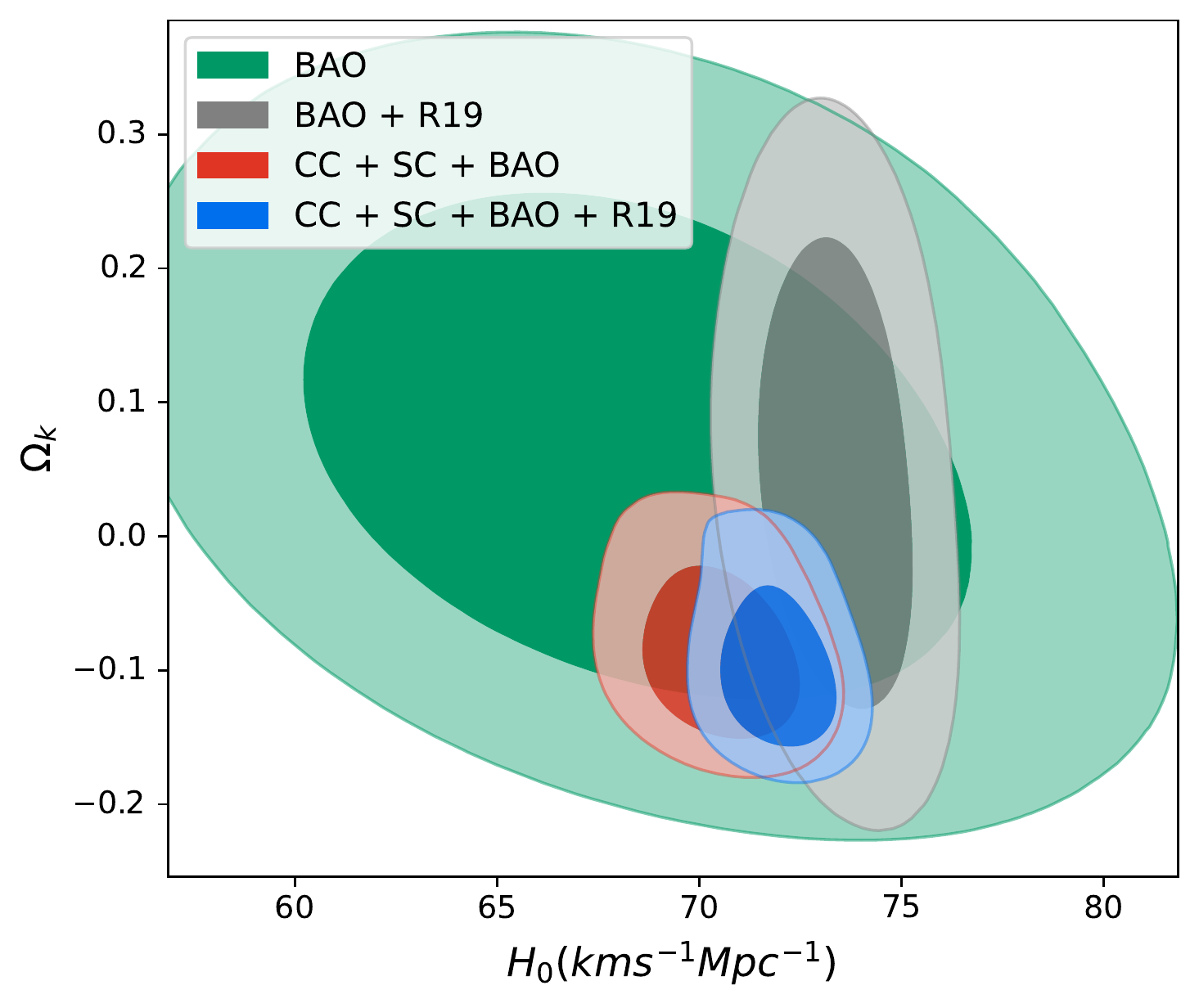}
  \caption{2d contour plot of $\Omega_k$ vs $H_0$ for the datasets BAO, BAO + R19, CC+SC+BAO and CC+SC+BAO+R19}%
  \label{fig2}
\end{figure}

Finally, we would like to discuss the results we obtained about the spacial curvature energy density. As already mentioned we get $\Omega_k = -0.076 \pm 0.012$ which excludes flat universe at 68\% CL and points to a closed Universe ($k=1$). Due to the rather small prior used to obtain this number, we now repeat the inference with a larger prior $\Omega_k \in [-0.3,0.3]$. The results can be seen on Fig. 2. Once again, when calculating for the full dataset with or without the R19 prior, we obtain values excluding flat universe.  This has been reported already, for example by the Planck 2018 collaboration (\cite{Aghanim:2018eyx}) for CMB alone which found a preference for a closed universe at $3.4\sigma$. Also \cite{Li:2019qic}, using the CC, Pantheon, and BAO measurements concluded that negative  $\Omega_k$ also relieves the $H_0$ tension.

\begin{figure}[!htb]%
\centering
  \includegraphics[width=0.5\textwidth]{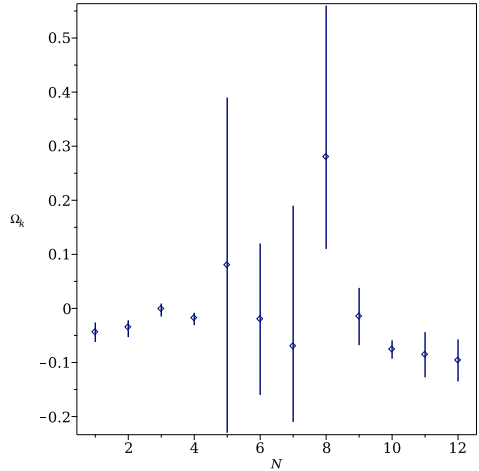}
  \caption{Comparison of different measurements of $\Omega_k$, see legend in text. Our points are $N=9-12$, where points with $N=11,12$ are with larger prior for $\Omega_k$}%
  \label{fig3}
\end{figure}

The issue of a possibility for a deviation from a flat Universe is extremely important and has been discussed in a number of works. On Fig. \ref{fig3}, we plot some of the published in the literature results with respect to $\Omega_k$. The reference for the results are accordingly ($N=1..12$) data from: Planck18Plk, Planck18CamSpec, ACT+WMAP, ACT+Planck, CC+Quasars,CC+SN, BAO+BBN+H0LiCOW, BAO+BBN+CC, taken from \cite{DiValentino:2020srs} to which we add the values we measured with the following dataset: BAO+R19, BAO+CC+SC (same as BAO+CC+SC+R19) and the extended prior points: BAO+CC+SCl, BAO+CC+SC+R19l. One can see that our results seem to add to a mounting evidence that maybe we observe an effect related to a non-flat universe.

The results for the full dataset for the extended prior can be seen in Table \ref{tbl3}. We see now that the reported earlier in Ref. \cite{Benisty:2020otr} lower matter energy density is no longer an issue as we get $\Omega_m\sim 0.326$, also we see that now $H_0$ is a bit larger, thus pointing to further alleviation of the $H_0$-tension due to a non-zero spatial curvature. Critically, we see that again $\Omega_k=0$ is excluded from the 68\% CL for both models. Thus we can claim this effect persists with the increase of the prior and it is a feature of the dataset.

On a side note, we mention the result we obtained in Ref. \cite{Benisty:2020otr} with respect to the $w$CDM model. For BAO + R19 we get $w<-1$ for  ($w=-1.067 \pm0.065$), while the full datasets seem to tend to $w\ge -1$ ($w=-0.989 \pm0.049$). Since $w=-1$ is included in the error, the results are essentially consistent with a cosmological constant. 

To compare the extended models to $\Lambda$CDM we used well known statistical measures, i.e. the Akaike information criteria (AIC) (\cite{AIC1,Liddle:2007fy,Anagnostopoulos:2019miu})\footnote{The definition of AIC we use is $AIC=\chi_{min}^2+2p$, where $p$ is the number of parameters of the model}. We find that the $\Lambda$CDM model remains the best fit to the data with difference of $2.3$ and $7.6$ AIC units for the $w$CDM and the $\Omega_k$CDM respectively. One can see that $w$CDM has a little bit more support than $\Omega_k$CDM but also as we mentioned, $w=-1$ enters into the the 68\% CL of $w$ for the full dataset which may explain its closeness to $\Lambda$CDM in terms of AIC. Our results are consistent with  the eBoss collaboration official results (\cite{Alam:2020sor}) . Interestingly, we see that in our case $w\ge-1$ for the full dataset, which differs from the estimations in some of the different cases  considered in \cite{Alam:2020sor}. Our dataset differs from theirs by the inclusion of the quasars and the GRB data and the exclusion of Planck points, so this may point to a local universe effect.

\section{Conclusion}
We use a set of 17 uncorrelated BAO points to infer the cosmological parameters for 3 different models: $\Lambda$CDM, $\Omega_k$CDM and $w$CDM. We find that by choosing the sound horizon at drag epoch as an independent parameter and adding additional datasets such as SNe, GRB and quasars, we are able to break the degeneracy between $H_0$ and $r_d$ and to constrain the cosmological parameters for the different models. The Hubble parameter obtained from the full dataset is very close to the TGRB measurement, while the one with a R19 prior is close to the local universe measurement.  We show that we are able to alleviate the $H_0$-tension to certain degree but not entirely. Another interesting result is the prediction of a non-flat but closed Universe at 68 \% CL, which has been confirmed with an increased prior and seems to add to the mounting evidences there may be a deviation from flatness.

\section*{Acknowledgments}

D.S. is thankful to Bulgarian National Science Fund for support via research grants KP-06-N38/11. We have received partial support from European COST actions CA18108.

\bibliographystyle{ws-procs961x669}
\bibliography{MG16_Staicova}

\begin{appendix}

\begin{table}
\tbl{The priors used to obtain each cosmological parameter:}
{\begin{tabular}{@{}cccc@{}}
\toprule
Parameter & Prior & Average Value ${}^1$& Average Value R19 ${}^2$\\
\hline
$\Omega_m$ & $ [0.;1.]$ & $0.26\pm 0.017$ & $0.26\pm 0.016$\\
$\Omega_\Lambda$ & $[0.;1 - \Omega_{m}]$ & $0.749\pm 0.013$ & $0.751\pm 0.013$ \\
$H_0$ & $[50;100]$ & $70.19\pm 1.11$ & -\\
$H_0^{R19}$ & $74.03 \pm 1.42$ & - & $71.68\pm 0.9$\\
$r_d/r_{d,fid}$ & $ [0.9,1.1]$ &$0.996\pm 0.019$& $0.97\pm 0.013$\\
$r_d$ & $[100;200]$ & $145.8\pm 2.37$ &$143.3\pm1.9$\\
$w$ & $ [-1.25;-0.75]$ & $-0.989\pm 0.049$  & $-0.989\pm 0.049$ \\
$\Omega_k$ & $ [-0.1;0.1]$& $-0.076 \pm 0.017$ & $-0.076\pm 0.012$
\\\botrule
\end{tabular}}
\begin{tabnote}
 ${}^1$ average value for the parameter under the flat prior for $H_0$, ${}^2$ average value for the parameter under the Gaussian R19 prior for $H_0$. $H_0$ is in $(km/s)/Mpc$, $r_d$ in Mpc.  \\
\end{tabnote}\label{tbl1}
\end{table}

  \begin{table}
\tbl{Numbers used to draw Fig 1:}
{\begin{tabular}{@{}ccccc@{}}
\toprule
Mission & Reference & $H_0$ (in km/s/Mpc) & $r_d$ (in Mpc)& $\Omega_m$\\
\hline
TDCosmo IV & \cite{Birrer:2020tax} & $74.5^{+5.6}_{-6.1} $ & - & - \\
TDCOSMO+SLACS & \cite{Birrer:2020tax} & $67.4^{+4.1}_{-3.2}$& - & - \\
\hline
H0LiCOW XIII & \cite{Wong:2019kwg} &$73.3^{+1.7}_{-1.8}$ & - & - \\
\hline
LMC DEBs  &\cite{Riess:2019cxk} &$74.22\pm 1.82$& - & - \\
LMC DEBs and NGC 4258 and Milky Way: &\cite{Riess:2019cxk} & $74.03\pm 1.42$& - & -\\
\hline
Planck 2018 \\
TT+lowE & \cite{Aghanim:2018eyx} & $66.88\pm 0.92$ & $147.21\pm0.48$ &$0.321\pm0.013$\\
TE+lowE & \cite{Aghanim:2018eyx} &  $68.44 \pm 0.91$ & $147.59 \pm 0.49$ & $0.301 \pm 0.012$\\
EE+lowE & \cite{Aghanim:2018eyx} &  $69.9 \pm 2.7$, & $146.46  \pm 0.70$ & $0.289^{+0.026}_{-0.033}$\\
TT,TE,EE+lowE & \cite{Aghanim:2018eyx} &  $67.27\pm 0.60$ & $147.05\pm0.30$ & $0.3166  \pm 0.0084$ \\
TT,TE,EE+lowE+lensing & \cite{Aghanim:2018eyx} & $67.36\pm 0.54$ & $147.09 \pm 0.26$ & $0.3153 \pm 0.0073$ \\
TT,TE,EE+lowE+lensing+BAO& \cite{Aghanim:2018eyx} &  $67.66\pm 0.42$ & $147.21 \pm 0.23$ & $0.3111 \pm 0.0056$ \\
\hline
eBOSS\\
BAO+BBN &\cite{eBOSS:2020yzd}& $67.35\pm 0.97$ & $149.3 \pm 2.8$ & $0.314\pm 0.008$\\ 
BAO and distance ladder & \cite{eBOSS:2020yzd}&$73.7\pm1.1$ & $135.9 \pm 3.2$ & - \\  
\hline
BAO + SDSS (BAO) + BBN & \cite{Nunes:2020uex} &$68.32_{-1.1}^{+0.98}$ & $151.9_{-2.8}^{+3}$ & $0.27^{+0.015}_{-0.016}$\\
\hline
BAO +BBN+H0LiCOW & \cite{Nunes:2020uex} & $74.88 \pm 1.95.1$ & $144.1\pm 5.3$ & $0.2763 \pm 0.027$\\
BAO +BBN+CC& \cite{Nunes:2020uex} &$72.06^{+1.2}_{-1.3}$ & $150.4^{+2.7}_{ -3.3}$ & $0.2515 \pm 0.016$
\\\botrule
\end{tabular}}
\label{tbl2}
\end{table}

\begin{table}
\tbl{The results for the $\Omega_k$CDM model with the increased prior on $\Omega_k\in [-0.3,0.3]$:}
{\begin{tabular}{@{}ccc@{}}
\toprule
Dataset & Parameter & Value \\
\hline
CC+SC+BAO & $H_0 (km s^{-1} Mpc^{-1})$ & $70.48 \pm 1.23$ \\
& $\Omega_k$ & $-0.086 \pm 0.042$ \\
& $\Omega_m$  & $0.326 \pm 0.027$ \\
& $\Omega_\Lambda$ & $0.766 \pm 0.029$\\
& $r_d (Mpc)$ & $145.961 \pm 2.676$ \\
& $rat$ & $0.984 \pm 0.016$ \\
\hline
CC+SC+BAO+R19 & $H_0 (km s^{-1} Mpc^{-1})$ & $71.91 \pm 0.87$ \\
& $\Omega_k$ & $-0.096 \pm 0.039$ \\
&$\Omega_m$ & $0.327 \pm 0.026$ \\
&$\Omega_\Lambda$ & $0.776 \pm 0.024$ \\
& $r_d$ (Mpc) & $143.452 \pm 1.948$ \\
& $rat$ & $0.967 \pm 0.013$
\\\botrule
\end{tabular}}
\label{tbl3}
\end{table}

\end{appendix}

\end{document}